\documentclass[10pt, conference, compsocconf]{IEEEtran}
\usepackage{graphicx}
\graphicspath{ {images/} }

\ifCLASSINFOpdf
\else
\fi
\begin{document}
%
\title{ColVis: Collaborative Visualization Design Workshops for Diverse User Groups}


\author{\IEEEauthorblockN{Damla \c{C}ay}
\IEEEauthorblockA{\textit{KUAR} \\
\textit{Ko\c{c} University}\\
\textit{Istanbul, Turkey} \\
\textit{Email: dcay13@ku.edu.tr}}

\and

\IEEEauthorblockN{Till Nagel}
\IEEEauthorblockA{\textit{Faculty of Computer Science} \\
\textit{Mannheim University of Applied Sciences}\\
\textit{Mannheim, Germany} \\
\textit{Email: t.nagel@hs-mannheim.de}}

\and

\IEEEauthorblockN{As{\i}m Evren Yanta\c{c}}
\IEEEauthorblockA{\textit{KARMA, KUAR} \\
\textit{Ko\c{c} University}\\
\textit{Istanbul, Turkey} \\
\textit{Email: eyantac@ku.edu.tr}}
}


%


\maketitle

\begin{abstract}
Understanding different types of users’ needs can even be more critical in today’s data visualization field, as exploratory visualizations for novice users are becoming more widespread with an increasing amount of data sources. The complexity of data-driven projects requires input from including interdisciplinary expert and novice users. Our workshop framework helps taking design decisions collaboratively with experts and novice users, on different levels such as outlining users and goals, identifying tasks, structuring data, and creating data visualization ideas. We conducted workshops for two different data visualization projects. For each project, we conducted a workshop with project stakeholders who are domain experts, then a second workshop with novice users. We collected feedback from participants and used critical reflection on the process. Later on, we created recommendations on how this workshop structure can be used by others. Our main contributions are, (1) the workshop framework for designing data visualizations, (2) describing the outcomes and lessons learned from multiple workshops.
\end{abstract}

\begin{IEEEkeywords}
User centered design, Participatory design, Information visualization
\end{IEEEkeywords}

%
\IEEEpeerreviewmaketitle

\section{Introduction}
Today, the increasing amount of data sources like sensors and activity trackers make data visualization an everyday life topic for many types of users. Thus designers are creating compelling visualizations that would be both useful and interesting for diverse groups of users including both novices and domain experts. Domain experts are defined as researchers who perform complex data analyses using visualization tools \cite{kirby2013visualization}, whereas novice users are new or inexperienced to certain tasks\cite{fisher1991defining}. Designing visualizations that target different types of users, can become challenging when there are unclear project goals, ambiguous tasks, unstructured data, or many different stakeholders. 

Collaborative user-centered design methods can be beneficial to understand the above mentioned critical data visualization aspects\cite{heer2008creation, kirby2013visualization}. Collaborative methods have been used in the data visualization field, however current approaches often emphasize data first, so user needs can be neglected\cite{huron2017let,huron2016using}. On the other hand, design thinking is a collaborative approach that focuses on users first to tackle complex problems\cite{brown2010design}. Typically, the two main objectives of a design thinking workshop are identifying the problem space and the solution space. In the first, participants understand the users and reframe the problem. During the second phase, participants ideate on solutions, build and test prototypes. 

Aligned with this approach, we suggest a workshop framework where the problem space deals with the goals, questions, and tasks of the users, while the solution space considers the data and visualization possibilities. In the problem space phase, participants define the target users, identify project goals, collect and prioritize questions and tasks according to these goals. If there are ambiguous terms in questions that are not directly linked to the data at hand, they define proxies to resolve the ambiguity. In the solution space phase, participants link the questions and tasks to data. They identify the necessary data, discuss available data. Lastly, participants have a general discussion on what kind of information is needed at what kind of granularity to support specific tasks. 
 
In this paper, we describe the workshop we conceived and implemented. We organized four workshops, two workshops each for two projects from different domains. The first workshop of each project had expert users, and the second had novice users. Then we collected feedback from the participants, and critically reflected upon our experiences to identify challenges and opportunities. We report recommendations for applying collaborative design workshops for designing visualizations for diverse user groups. Collaborative design for data visualization is a challenging subject considering the level of expertise required for building data visualizations. Our findings indicate that these workshops can enable collaboratively taking design decisions that reflect the values of different stakeholders.

\section{Related Literature}

\subsection{Participatory and User-Centered Design}
Design knowledge is the knowledge embedded in the design of an artifact or service \cite{frauenberger2015pursuit}. Creating design knowledge through end-user participation has been an important aspect of participatory design\cite{brandt2006designing, ehn2008participation}. As Schön \cite{schon2017reflective} defines, participatory design is the process of mutual understanding, investigating, reflecting between participants where designers learn the realities of users and users articulate their aims. Useful methods for participatory design include workshops, cultural probes \cite{gaver1999design}, ethnography, and cooperative prototyping \cite{simonsen2012routledge}.

Among these methods, the workshop technique has been widely used for human-computer interaction (HCI) research. Different workshop methods and tools are developed for different aims and settings. Organizations like IDEO and Stanford d.school have been successfully implementing user-centered design thinking workshops for business solutions and social innovation \cite{brown2010design, plattner2009design}. These workshops typically include hands-on divergent and convergent activities with users to explore and prioritize possibilities \cite{osborn2012applied}.

Collaborative and user-centered design methods are rapidly gaining popularity among data visualization researchers and practitioners as well\cite{munk2019data, arcia2015sometimes, gencc2015participatory, ccay2016learning}. He and Adar \cite{he2016v} express that design thinking could be useful for information design cases because of the wickedness of the data visualization design studies. Wicked problems are without definitive limits or conditions to the design problem \cite{buchanan1992wicked}, and visualization design studies can be defined as wicked problems due to the iterative nature of the design study research, as elaborated by Meyer and Dykes \cite{meyer2019criteria}. With this perspective, we believe pursuing a design thinking approach with divergent and convergent activities would be useful for data visualization cases.

\subsection{Frameworks for Data Visualization Design}
Existing data visualization design frameworks formulate steps to take when designing interactive data visualizations. Munzner \cite{munzner2009nested} proposed the nested model. The model identifies four nested decision-making levels which are; (1) domain problem characterization, (2) data/operation abstraction design, (3) encoding /interaction technique design and (4) algorithm design. Another well-established framework from Sedlmair et al. \cite{sedlmair2012design} describes nine visualization design stages as learn, winnow, cast, discover, design, implement, deploy, reflect, and write. 

Consistently with these frameworks in the literature, the framework that we suggest starts with defining the problem space first and then focusing on data, while different stages of our workshop framework generate information that supports both of the above-mentioned taxonomies by covering; domain problem definition, data/operation abstraction design and encoding/interaction technique design of the 4 stages and understand, ideate, winnow, cast and discover stages of the nine-stage framework.

\subsection{Co-design methods for Data Visualization}
User-centered and collaborative methods are getting popularized. Koh et al. \cite{koh2011developing} propose a user-centered visualization design approach where the process starts with familiarizing users with visualization methods through collaborative activities. Collaboration for data visualization used to take place between visualization researchers and other types of researchers. When fields like economics, business, and humanities started to use data visualizations increasingly, they were included in participation as domain experts \cite{kirby2013visualization}. Today, data visualizations are not only used as analysis tools for researchers and experts but also for data exploration by novice users\cite{nagel2014touching, arcia2015sometimes}. This requires the collaboration sphere to expand to novices\cite{mckenna2017worksheets}. Heer et al. \cite{heer2008creation} characterize the visualization user base as expert, savvy or novice users. They identify a new research goal of supporting novice users to specify their needs for a visualization. Our workshop framework enables these different types of users to specify and prioritize their data visualization needs.

As opposed to sequential visualization design frameworks, Wood et al. \cite{wood2014moving} perform simultaneous user studies with different user types for a visualization case study. The authors state that this technique enables them to gain rich insights to guide the visualization design process. They use various methods like public releases, talks, workshops, stakeholder meetings to gain insights through a three years long period. Hall et al. \cite{hall2019design} learn from users through immersive exchanges between visualization researchers and domain experts. Our approach is similar in the sense of gaining rich insights from different types of users through different activities over time. However, we aim to initiate this process in a shorter period.

Kerzner et al. \cite{kerzner2018framework} define guidelines for workshops of data visualization opportunities. Authors argue for participants to adapt to a visualization mindset and recommend different design activities for different purposes, where we focus on a systematic yet flexible structure which allows faster planning and execution. Differently, our approach starts with solely focusing on the problems and need, then focusing on visualization solutions.  Additionally, our workshop framework enables exploring real data as well, and find solutions with data in mind. Our work builds upon the existing literature of collaborative practices in visualization and adapts it to a setting where different user types including non-experts can take part in the data visualization process.


\section{Collaborative Visualization Workshops}
Based on the challenges we experienced during earlier studies on building visualization tools with interdisciplinary teams, we opted for collaborative design workshops. We started working on a workshop structure to create a generalizable framework to use as a guideline for planning and executing workshops, reporting outputs to gather problems and ideas for a specific data-driven real-world context for diverse stakeholders more richly and creatively. 

In this section, we explain the workshop framework we have conceived and refined iteratively. Later on, we report on the lessons learned through performing the workshops for visualization projects from different domains. We critically reflected upon our experiences during the workshops and analyzed feedback from the participants. After refining the framework using these insights, we observed how the workshop worked independently by collecting feedback from a moderator who used the workshop for their project.

In the following section, we will present four workshops conducted for two different projects. For both projects, we first conducted a workshop with project stakeholders who are domain experts, then a second with novice users. At least one visualization expert was present at the workshop and they moderated the workshops. The purpose of the visualization expert in all workshops was to translate the discussions into design decisions at the design phase. Finally, an additional workshop on a third visualization case was conducted by a moderator who is not a part of the team. 

\subsection{Projects and Participants} 
The first visualization project, “The City Walls” is a collaboration between Archaeology and Design departments that aims to create a geographical visualization of data related to the city walls of Istanbul. The data the archaeology team collected includes historical information about the city walls from primary historical sources, historical images and footage, architectural details of the walls, and a photo archive created by the team that includes detailed images of each gate, tower, and wall.

For the first workshop with experts, we invited all the project stakeholders. The workshop participants were 6 archaeologists, 1 photographer responsible for creating the photo archive, 2 designers and a developer (3 female, 7 male). The workshop lasted for 5. 5 hours. For the second workshop with novice users, we announced to the network of a co-working space in Istanbul.  One game developer, one architect, one interior architect, and one visualization expert attended the workshop (1 female, 3 male). The workshop lasted for 2. 5 hours. 

The second visualization project called "Hope Archive", is a geospatial video archive about non-governmental organizations’ (NGO) activities. The project aims to make the NGO activities visible, establish spatial or contextual connections among different NGOs. The initial motivation for the video database started with the activities of the Düzce Hope Homes. These videos documented the struggles of the 1999 earthquake victims and the participatory process of redesigning and rebuilding a living space for them. The data related to the videos are actors, themes of the NGOs, and tools for the activity.

For the first workshop with experts, we invited all project stakeholders to the workshop. 8 NGO employees, 1 documentarist, 2 designers, 1 developer (6 female, 6 male). The workshop lasted for 6 hours. For the second workshop with novice users, we announced it to the network of a co-working space in Istanbul. One project coordinator, one service and UX designer, one community coordinator and one visualization expert (4 females) attended the workshop, and it lasted for 3 hours.

\begin{figure*}[tbp]
  \centering
  \hfill
  \includegraphics[width=0.85\linewidth]{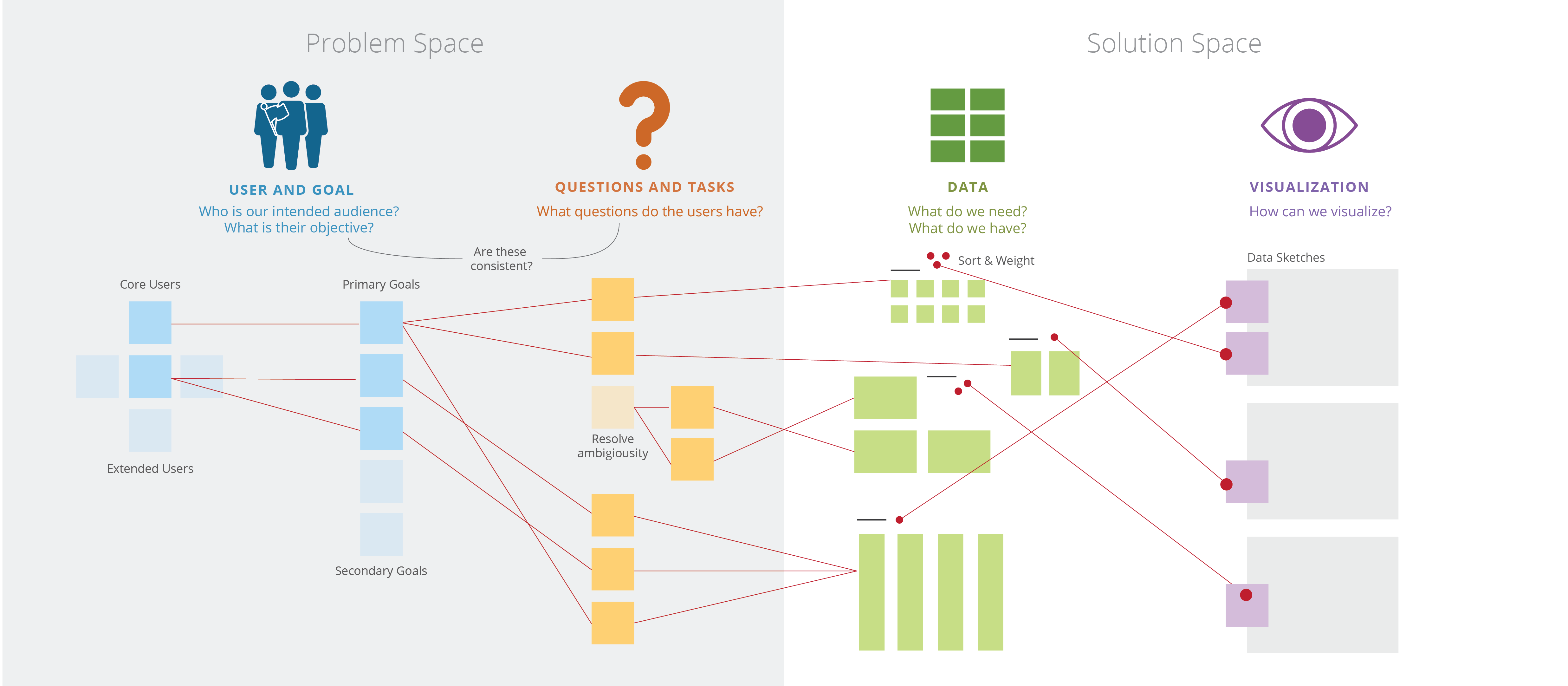}
  \hfill \mbox{}
  \caption{\label{fig:colvis_process.png}%
            The workshop has four phases :1. User and Goal, 2. Questions and Tasks, 3. Data, and 4. Visualization. The lines represent possible relations between consecutive phases. Detailed guide of the process and other materials can be achieved at \url{https://github.com/colvis2019/ColVis-Workshop} 
            }
\end{figure*}

\subsection{Procedure} 
In the workshops, we employed the general structure of a design thinking workshop where participants first define the problem space and then define the solution space through divergent and convergent activities.\newline 
The workshop has four phases as demonstrated on Figure~\ref{fig:colvis_process.png}: \newline
1.    User and Goal (Problem space) \newline
2.    Questions and Tasks (Problem space) \newline
3.    Data (Solution space) \newline
4.    Visualization (Solution space)  \newline

The User and Goal phase starts with an open discussion. Participants discuss and list the potential users of future visualization. Then, they elect the core and extended users of visualization using the dot voting method to prioritize user types (Figure~\ref{fig:colvis_process.png}, Define Core and Extended Users). Next, participants discuss and list the goals of the prioritized possible users, then vote for the most important and relevant goals. 

In the Questions and Tasks phase, participants discuss and list the questions to ask to the visualization, considering the goals they define in the previous phase\cite{lam2017bridging}. Participants this time, vote the questions that are the most relevant to the goals or interesting. Participants create one or more tasks out of each selected question. Then they identify ambiguous components (not directly addressable by the dataset). Participants define proxies until all tasks are actionable \cite{fisher2017making}. This workshop phase aims to form clear tasks from ambiguous questions.
 
After this, participants continue to data phase where the aim is to identify links with questions and data. If there is already collected data, participants explore the data set to identify the links between questions and data. If there is no or partially collected data, participants discuss which data is needed to answer the questions. In this phase, participants use methods like card sorting, affinity diagramming, mind mapping, and dot voting to organize and prioritize data. 

In the visualization phase, we present different visualization functions (Distribution/ Time / Compare / Geospatial / Part-to-Whole / Relationship) and interaction styles (selecting / filtering / brushing / hovering / highlighting).  Then we demonstrate examples of the explained concepts. After this, participants discuss which dimensions of data should be visualized.  Then they ideate on alternative visualization ideas. This activity can be conducted as a group or individual activity depending on the number of participants and participant preference. If it is conducted as an individual or small group activity, at the end of the workshop, the participants present and give feedback to the ideas. If it is conducted as a group activity with all participants of the workshop participants, it can be implemented as group ideation followed with a reflective discussion. There are two outputs of this phase that can inform the design of the visualization. Firstly, the visualization ideas that received positive feedback can inform abstraction and interaction design. Secondly, critique and discussions reveal the final design requirements. 

\subsection{Methodology} 
After the workshops, we collected the notes from the workshops, answers to the post-workshop open-ended questions, our notes from the oral feedback participants gave after the workshops and the critical reflections \cite{brookfield1998critically, meyer2018reflection} of the authors who participated as visualization experts at the workshops.

\section{Reflections about Workshops}
In this section, we briefly present an overview of each workshop's process. Then we present our critical reflections about the process and highlight the important points of the participant feedback. 

\subsection{W1. The City Walls Map with Experts} 
After a brief overview of the project's main goals and status, we started the workshop with users and goals. The participants prioritized novice citizens as the primary user type and “exploring the city walls” as the primary goal. In the next phase Questions and Tasks, the participants generated and prioritized questions. For the data phase, we printed out samples from the visual content and excel sheets before the workshop. During the data phase, first, the participants identified the links between the questions and existing data. Some questions required data other than the already collected data. These were also identified during the discussion. After this, participants sorted and prioritized the data samples using card sorting and dot voting methods. At the visualization phase, first, participants discussed the different visualization possibilities over examples. Then they created data sketches and the workshop finished with a group discussion about the results. 

Before the workshop, the project stakeholders had a vague definition of the project goal. They started collecting visual data with an archival motivation and wanted to create a geographical visualization from this archive. They stated that such a tool can be useful for remote researchers. However, during the workshop, the necessity of defining and prioritizing the user and goal made the stakeholders realize that they were prioritizing citizens as users over researchers. During the design process before the workshop, the project stakeholders and the visualization team had several meetings. These meetings included different combinations of stakeholders at once, due to availability. Discussions during the workshop revealed that different stakeholders had different visions of and expectations from the visualization tool. The workshop structure enabled them to create a unified goal.


Post-workshop feedback from the participants reflects that they were overall pleased with the workshop at the end. One participant stated, “At the beginning, I wasn't quite sure where it will all lead but I was impressed with the results we ended up with.” One participant expressed the need for more breaks. Some participants felt like one stakeholder dominated the discussions for some phases of the workshop.

\subsection{W2. The City Walls Map with Novice Users}

In the second workshop of the same project, the participants identified students as the prioritized user type and, “exploration and research” as the primary goal.  In the Questions and Tasks phase, the questions participants generated were related to the main entry paths to the city, and the modern socio-cultural surrounding of the walls. Even though some questions that were generated in the workshop were similar to the ones from the expert workshop, the prioritization differed. Novice users focused more on gates than other architectural elements like walls or towers. They also prioritized contemporary information like the communities lived and still lives around the city walls. At the data phase, the group was presented the same material from the expert workshop, including architectural data, historical data, and visual material. However, they had trouble linking the existing data to some of the questions and proposed new data types instead. At the visualization phase, the participants proposed visualization ideas for different levels of detail, as a group.
 
Participants felt that the workshop had a casual and relaxed environment. Even though we explained the overarching aim of the project at the beginning and presented the collected data, some participants expressed that they felt uninformed about the project. One participant found the discussions too free-form. One participant found the discussions to be too abstract, another enjoyed the dialog and discussion itself. Several participants expressed that the workshop's interdisciplinary nature helped to create fruitful discussions. One of them found, “The difference of participants in terms of background and discipline enables each other to see new aspects and create a cohesive contribution.” 

Several participants felt like the collaboration took place in the form of building upon each other's ideas. One participant said, “The act of sharing all our individual ideas on topics was itself the collaboration.” Another participant found the use of post-its enabled the discussion to be more visible and this helped to trigger their participation.

\subsection{W3. Hope Archive with Experts}

Following the same structure, participants prioritized the Researcher/Student using the visualization for research. The second user type was NGOs, using the visualization to learn best practices and networking. The third user type was journalists, using the visualization to find stories. At this point, one participant opened the discussion of content creation around the questions like, if the platform will be open to the public, will it be moderated or unmoderated, or will it be a closed platform where people can apply with their content.
 
In the Questions and Tasks phase, the prioritized questions were, what type of activities are NGO’s engaged in? Where do these activities take place? What are the methods they use? What are the NGO activities with a higher impact? After the questions are set, the data dimensions related to the questions were, video, story, location, actors, theme, method, the amount of content, latest content. The visualization decisions included having a simpler base map, improving the visual connection between the map view and list view on the existing tool, functional suggestions like connecting YouTube channels to the website and automating the video upload process. Other suggestions were related to fixing the usability issues of the existing tool.

After the workshop, we identified two important points while critically reflecting on the process. Firstly, the final discussion did not involve visualization solutions according to the identified data types. They were functional enhancements for the existing tool. Secondly, during the workshop, one participant initiated discussions repeatedly on who will produce the content of the platform and how. This repetition caused a loss of focus and prevented the discussion from moving forward at times. 
The qualitative feedback we collected after the workshop included recurring themes. Some participants expressed that the workshop helped them clarify goals and discuss the tool thoroughly. Additional positive comments stated that the workshop created awareness of the problems and awareness of the necessity to use more user-centered methods. In terms of teamwork, two participants stated that the workshop was more like a place to share individual ideas rather than teamwork. The negative comments were related to the repetition of discussions. Some participants felt like the workshop structure was unsystematic, the discussions were too broad and there was no clear result at the end of the workshop.

\subsection{W4. Hope Archive with Novice Users}

The workshop started with an introduction where the moderator explained the goals and motivations of the project. The prioritized user types were students/academics, NGOs/collectives, and local governments. The identified goals were: researching for students/researchers, archiving their projects and networking for NGOs/collectives, finding project stakeholders for local governments. The questions generated were, who are the people doing similar work to our NGO? What type of methodologies they use? What has been done on a specific topic? When was it done? Does anyone have data that I can use? Related to the questions, the data types identified by participants were the location of the NGO, topic, methods, photos, videos, publications and date of each activity and references. At the visualization phase, for the overview, participants proposed a network visualization where users can see the links between NGOs and topics. At this level, they wanted to see the NGO name, topic, stakeholders, starting and last active dates. 

Our critical reflections on the workshop process include two important points. Firstly, the discussions were more clear and fruitful than the expert workshop. Outcomes of the workshop were more suited to guide the visualization design. Secondly, to make the goal and the content clear, we showed the existing prototype. However, this limited the participants, to the point where they can only identify the usability issues. After they are reminded to think freely without limiting themselves with the existing tool, they started to ideate.

From the post-workshop questions, one common positive comment was about the flow of the discussion and the moderator’s guidance. One participant stated that the moderator successfully guided the discussion when it was out of focus and another commented on the moderator synthesized and framed the outcomes effectively. Another positive aspect was about visualization awareness. One participant stated that the workshop enabled them to think about their data-related projects more clearly. Others were glad to be aware of a local project that might be of interest. Overall, they felt like it was successful and enjoyable teamwork. On the other hand, some participants commented on the negative impact of seeing the existing tool. One participant stated that it limited the discussion. Another negative point that one participant felt not informed enough at the beginning of the workshop.


\subsection{Case Study:  Sonic Memories}

After four workshops moderated by visualization experts who also author this paper, we wanted to have an additional workshop with a non-team member as a moderator, to test and improve the workshop framework. We prepared a detailed moderator’s guide that included the phases and steps to conduct the workshop independently. One researcher who was starting with a new data visualization project used the framework, whose project deals with visualizing personal memories related to city sounds. After they conducted the workshop, we interviewed the moderator and two workshop participants.

Overall, the researcher found the workshop to be useful in terms of identifying and justifying data needs.
The moderator stated, "The data phase was useful. I collected sample data from the participants, everyone wrote memories about city sounds. Then we extracted data dimensions from those. The dimensions were similar to what I had in mind before the workshop. So my ideas were supported in this phase. There were additional ideas about the functionality, which I haven't thought before."

Generating questions that are related to the prioritized user goals was a challenge. The moderator said, "Some questions generated in the questions and tasks phase were not questions about interacting with data. I had to intervene and re-direct a lot here.“ Additionally, one participant said, “Questions and Tasks phase was good but it wasn't clear which questions relate to which goals of which user type. We tend to forget about the user in this phase. We generated a lot of questions and some of them weren't related to a defined goal.” In addition to generating questions, prioritizing them using dot voting was also unclear and challenging. One participant stated, “When selecting questions with dot voting, I observed that people tended to select the ones that are easy to understand rather than interesting ones.” Similarly, another participant said, "The selection process of questions was hard. I wasn't clear on the selection criteria. We could have selected the wild questions but we didn't. 

The last important point both mentioned by the moderator and a participant was about the sketching part of the visualization phase. The moderator stated that the participants who weren't designers struggled when sketching. The moderator stated, "Maybe they can communicate their ideas differently.“

\subsection{Design and Development of the Hope Archive and the City Walls projects}

 \begin{figure*}[tbp]
  \centering
  \hfill
  \includegraphics[width=1\linewidth]{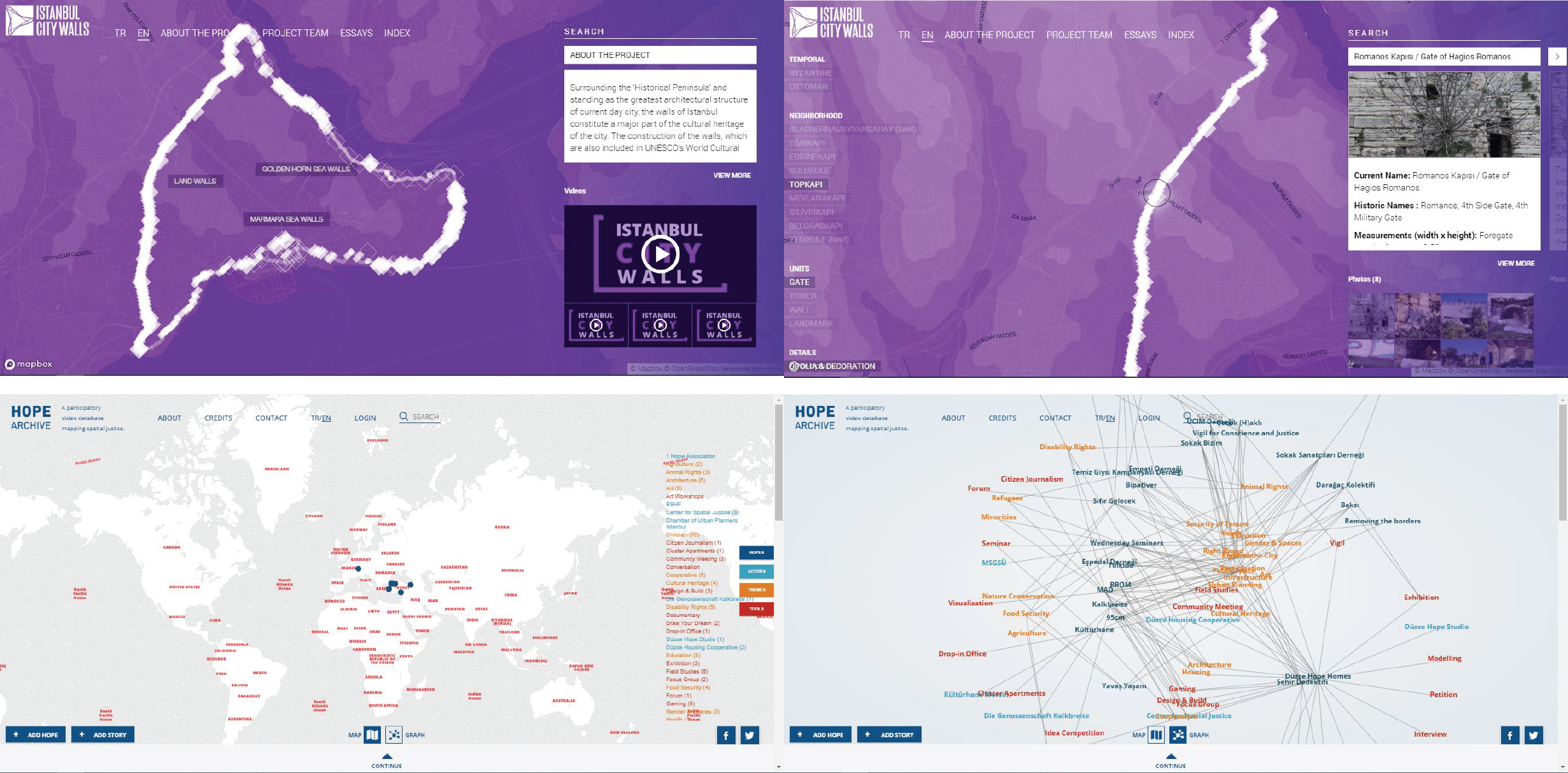}
  \hfill \mbox{}
  \caption{\label{fig:prototypes-01.png}%
           Hope Archive and The City Walls projects are realized and online. The links are provided in the supplementary files at   \url{https://github.com/colvis2019/ColVis-Workshop} }
\end{figure*}

Even though the aim of this paper is not to discuss the designed tools extensively, we would like to briefly give an overview of the tools. Both projects were designed and developed using the decisions from the workshops(Figure~\ref{fig:prototypes-01.png}). On the City Walls project's main page, information related to gates, towers and walls are visualized. Glyph for gates are bigger because they were stated as more important at both workshops (Figure 3, top left). When clicked, tags appear on the left side of the screen as suggested in the expert workshop and more detailed information about the unit is presented (Figure 3, top right). At the Hope Archive's main page, the information is presented geographically as the experts suggested (Figure 3, bottom left). Same data can be viewed in a node-link diagram as the novices suggested (Figure 3, bottom right).

\section{Discussion}
\subsection{Maintaining the focus for informing visualization design}
One of the biggest challenges during the workshops was keeping the focus on designing visualization and guiding the discussions in a way that will create useful outcomes for designing visualizations. Here we share the patterns we identified from our critical reflections and post-workshop feedback and recommendations to overcome problems that can occur.

In every workshop session, discussions shifted to subjects that were not directly about the visualization itself.  We observed these shifts were longer and deeper in expert workshops. It was harder to focus back on the visualization. Even though there may be discussion around the topic of interest, the main focus should be on the visualization. Another problem we encountered several times, was discussion shifting to a topic that is not directly related to the workshop phase. For instance, repeated discussions on the data collection method on every phase of the Hope Archive expert workshop. Our approach depends on starting with the user in mind, then moving towards questions and finding links between those questions and data. Each phase creates the outcome of the next one. This structure makes it important to focusing only on one phase at a time. To overcome these problems, we recommend moderators to selectively take notes by only having related keywords noted, and bring attention to the current workshop phase as necessary. 

For both expert workshops, there were long and insistent discussions about project-related, but not visualization related topics. These long discussions in each workshop were dominated by one participant, who were both project stakeholders. Aside from elongating the workshop period, this also affected other participants negatively. Dominant participants were mentioned negatively by other participants in the post-workshop survey. One participant even suggested that the moderator should decide who will talk when. To overcome this challenge, workshops can be divided to have up to 5 participants, as small workshops enable everyone to take part more comfortably. Another solution during the workshop could be, having a quick round around the table, asking everyone their individual idea, and then making a short, concluding group discussion. 

During the novice workshops, we showed a work-in-progress version of the visualization. Our aim was not to influence their visualization choices but to show the available data types. These unfinished visualizations caused distraction and unnecessary discussions about the usability problems of the interactive visualizations. During the novice workshop of the Hope Archive project, showing work-in-progress caused the divergent phases to be more limited. After the workshop, participants stated that the discussions were more productive after the moderator reminded them to think as if the work-in-progress did not exist. Since the workshop aims to reveal design requirements, we recommend not showing any work-in-progress material.

We arranged the workshop set up in a way that outcomes of the previous phase were visible either on a wall or table. However, during the case study, some participants and the moderator mentioned they had a hard time with the questions and tasks phase and some questions were not directly related to the prioritized user type and goal. We recommend visually and orally highlighting the prioritized outcomes of each phase, and intervene every time an unrelated input occurs, remind the participants the overall goal and the process of the workshop. 

We wanted to include the dot voting method when prioritization is needed as it is commonly used in design thinking workshops as a quick way to understand the group tendencies. When implementing this method for our data visualization workshops, sometimes if fell short for our need for prioritizing with important criteria in mind. These criteria were about how relevant, important, interesting or feasible something is. For the Sonic Memories case study, some participants stated that voting created confusion during the questions and tasks phase as they were not aware of why they voted, and each participant was voting for a different reason. Instead of using the same element (dot) as feedback, we recommend separating vote types visually, by either color-coding or writing the feedback types on the votes. 

\subsection{Nature of participation differs for designing data visualization}
Even though the participants of the two workshops were different people with entirely different levels of domain knowledge and involvement in the project, the groups generated similar questions during the Questions and Tasks phase for the City Walls project. On the other hand, experts and novices prioritized different questions. This reveals that our workshop framework, when applied to different groups of users, can be a way to understand the most important tasks for a visualization. Seeing different prioritizations can reveal different design requirements for different user types.

During the visualization phase, some participants refrained from creating sketches. This can be common among participants who are not from a design background. Design thinking workshops have special activities to encourage people to sketch. However, this might be hard to apply because of the time limit, and also unnecessary since the ultimate goal is to make design decisions that are based on needs and data. One solution might be creating collages in this phase \cite{chen2014exploring, bludau2017}, or having pre-made visual examples of basic visualization methods as sheets or cards for participants to communicate their design decisions easier. 

The City Walls project had more complex data types compared to the Hope project. Regarding the data complexity, the data phase of the City Walls expert workshop took the longest. Besides, experts had an easier time sorting data since they have expertise on the subject. On the other hand, both expert and novice users can identify interesting data types for projects aimed at diverse user groups. Our reflections on the process and feedbacks indicate that the depth of data and participant type affect the process, and the workshop should be applied considering these differences.
\subsection{Challenges of organizing and moderating a data visualization workshop}

Every data-visualization project has its unique challenges related to the data itself. Data might be unavailable, missing, unclear, or complex. We envisioned the workshop to work effectively with different amounts of existing data. If the project does not have any data, the data phase can be used to identify the needed data types and how they can be achieved. If there are data, but the project stakeholders are open to suggestions, then a similar discussion on data types can be followed by browsing existing data, sorting and prioritizing and finally identifying links between questions and data. At the Sonic Memories workshop’s data phase, the moderator who was also the project owner decided to collect sample data by asking participants to write a memory about a place. Then they were able to identify the data dimensions that the memories included and continued the rest of the data phase using these samples. After the workshop, the moderator stated that data phase was very useful for the project. When applying the workshop, we recommend adjusting the data phase according to the needs and circumstances of the project.

The space that the workshop happens in is an important element that affects the nature of participation. Since the available options to host a workshop might be limited, we envisioned the workshop to be applicable in a variety of spaces. However, there are two essential elements. The first one is the visibility of generated and prioritized keywords and how they relate to other phases. The second one is having enough space to display data and perform hands-on activities. A table or wall can be used for these purposes. In small workshops with up to 4 participants, a small table can be suitable to arrange post-its and data since everyone will be able to see and reach the material. However, a bigger workshop might require an empty wall, and enough space in front of the wall to place and organize the post-its. Additionally, the table can be used to organize data and create visualization ideas. Space should be considered along with the number of participants. Overcrowded spaces with more participants than the table can afford, can hinder hands-on participation.

\section{Conclusion}

We presented ColVis Workshop Toolkit, that enables creating human-centered data visualization solutions collaboratively with diverse user groups like novice and expert users. We designed this workshop to include users early into the data visualization process starting from defining users and goals, identifying and prioritizing tasks, identifying existing and needed data, and creating data visualization ideas according to the defined requirements. We applied the workshop framework to two projects, two workshops for each project, one with expert users and the other one with novice users as participants. Additionally, an external researcher implemented the workshop for their project. Based on our critical reflections and qualitative feedback of participants and an external researcher, we find that ColVis workshop structure provides data visualization design directions on different levels, in a user-centered way. We provide the recommendations and the material and hope they can be used and developed further to enable deeper user participation in the data visualization field.

\bibliographystyle{./bibliography/IEEEtran}



\end{document}